\documentclass[twocolumn,showpacs,showkeys,preprintnumbers,amsmath,amssymb,floatfix,superscriptaddress]{revtex4}

\usepackage[dvips]{graphicx}% Include figure files
\usepackage{dcolumn}% Align table columns on decimal point
\usepackage{bm}% bold math
\usepackage{amsmath,amsthm,amssymb}
\usepackage{graphicx}
\def\beq{\begin{equation}}
\def\eeq{\end{equation}}

\newcommand{\affA}{%
    Van der Waals-Zeeman Institute, University of Amsterdam,\\
Valckenierstraat 65, 1018 XE Amsterdam, The Netherlands }

\newcommand{\affB}{%
   Hitachi San Jose Research Center, San Jose (CA), USA}

\begin{document}

\preprint{}

\title{A lattice of microtraps for ultracold atoms based on patterned magnetic films}

\author{R. Gerritsma}\affiliation{\affA}
\author{S. Whitlock}\affiliation{\affA}
\author{T. Fernholz}\affiliation{\affA}
\author{H. Schlatter}\affiliation{\affA}
\author{J.~A. Luigjes}\affiliation{\affA}
\author{J.~-U. Thiele}\affiliation{\affB}
\author{J.~B. Goedkoop}\affiliation{\affA}
\author{R.~J.~C. Spreeuw}\affiliation{\affA}
\email{spreeuw@science.uva.nl}
\affiliation{}\homepage{http://www.science.uva.nl/research/aplp/}

\date{\today}% It is always \today, today,

\begin{abstract}
We have realized a two dimensional permanent magnetic lattice of
Ioffe-Pritchard microtraps for ultracold atoms. The lattice is formed by a
single 300~nm magnetized layer of FePt, patterned using optical
lithography. Our magnetic lattice consists of more than 15000 tightly
confining microtraps with a density of 1250~traps/mm$^2$.  Simple
analytical approximations for the magnetic fields produced by the lattice
are used to derive relevant trap parameters. We load ultracold atoms into
at least 30 lattice sites at a distance of approximately 10~$\mu$m from
the film surface. The present result is an important first step towards
quantum information processing with neutral atoms in magnetic lattice
potentials.
\end{abstract}



\maketitle

\section{Introduction}

Lattice potentials for ultracold neutral atoms and Bose-Einstein
condensates have been used in a series of spectacular experiments. Optical
lattices for example have allowed for the observation of the
Mott-insulator to superfluid transition~\cite{GreManEss02}, studies of low
dimensional quantum gases~\cite{ParWidHan04,HadKruDal06} and the coherent
production of isolated and long lived cold molecules
\cite{RomBesBlo04,StoMorEss06,ThaWinDen06,OspOspBon06,VolSyaRem06}. In
another remarkable experiment large scale quantum entanglement
\cite{ManGreBlo03} has been achieved in an optical lattice, pointing the
way to future applications in quantum information processing.

Magnetic lattice potentials are a promising alternative to the widely used
optical lattices that rely on intense overlapping laser beams. For
instance, magnetic lattices produced using periodically magnetized films
on atom chips provide a high degree of design flexibility allowing
arbitrary trap geometries and lattice spacings~\cite{GhaKieHan06}. The
potentials are also magnetic state selective and only weak field seeking
atoms remain trapped, allowing for manipulation using radio frequency (rf)
fields - a powerful tool used extensively for forced evaporative cooling
and for rf spectroscopy~\cite{FerGerSpr07}. Furthermore, magnetic lattices
provide stable and tight confinement even with large separation between
sites. It should also be possible to incorporate on-chip detection and
manipulation schemes~\cite{TepLinVul06} to address individual sites.

Magnetic lattices may be most promising as a scalable quantum system, a
crucial ingredient for quantum information processing~\cite{Div00}.
Trapped single atoms can be used as qubits, with their internal Zeeman or
hyperfine states representing the qubit states $|0\rangle$ and
$|1\rangle$. Superpositions with a long coherence time have already been
demonstrated near a chip surface~\cite{TreHomRei04}, indicating that the
coupling to environmental fields is weak.

In this paper we describe the design, fabrication and loading of a
two-dimensional lattice of magnetic (Ioffe-Pritchard) microtraps for
ultracold neutral atoms. The microtraps are produced near the surface of a
patterned FePt film with perpendicular magnetization. Our magnetic lattice
design consists of 1250 traps/mm$^2$, with the expectation that this
density can be scaled up by two or three orders of magnitude.  We describe
the properties of the magnetic film and our fabrication of the structure.
Finally, we load a cloud of ultracold $^{87}$Rb atoms into this lattice,
demonstrating that our approach is a feasible one. The results open up a
range of new possibilities for future experiments using permanent magnetic
atom chips.

The paper is structured as follows. In section~\ref{sec:model} we present
the lattice geometry based on a single layer of magnetised material which
in combination with a uniform bias field produces a two-dimensional array
of magnetic microtraps.  We provide relatively simple analytical
expressions to approximate the magnetic field patterns. In
section~\ref{sec:fept} we describe the properties of the magnetic FePt
films and our procedures for fabrication and patterning of the films. In
section~\ref{secLoading} we show that ultracold $^{87}$Rb atoms can be
transferred to the lattice from a Z-shaped wire magnetic trap, to occupy
more than 30 lattice sites. Finally, we discuss some prospects for future
experiments and in particular focus on quantum information processing
using atom chips.

\section{A two-dimensional magnetic lattice potential}\label{sec:model}

\begin{figure}
\center
\includegraphics[width=7cm]{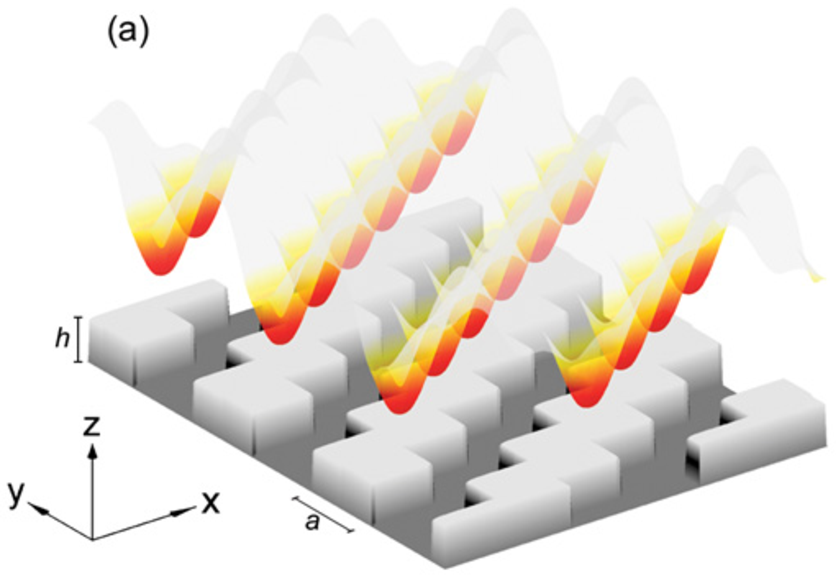}
\includegraphics[width=7cm]{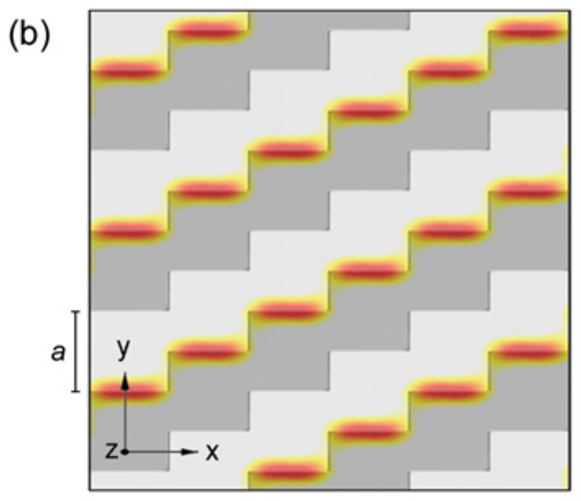}
\caption{(Color online) Schematic of a magnetic lattice potential for
ultracold atoms formed by a single layer of uniformly magnetized material.
The patterned FePt film (light gray) is magnetized along $z$. A uniform
bias field is applied along the $y-$direction to create an array of
Ioffe-Pritchard microtraps.  The perspective view (a) indicates the
potential energy above the film surface and the top view (b) shows the
position and orientation of the Ioffe-Pritchard microtraps as diffuse red
shaded areas; the parameter $a$ is a natural length scale of the geometry
and $h$ is the thickness of the magnetic film.} \label{figSchematic}
\end{figure}

We have considered several geometries for producing magnetic lattices and
have chosen a single design that produces an array of microtraps with
non-zero field minima, provides tight confinement to the atoms and can be
loaded in a straight-forward way.  Our permanent magnetic lattice design
is shown in Fig.~\ref{figSchematic}. The magnetization is oriented out of
the film plane (parallel to $z$ in Fig.~1b), and produces a field
analogous to an array of Z-wire magnetic traps with an equivalent current
of magnitude $I_{\rm{eq}}=M_0h$, where $M_0$ is the remanent magnetization
and $h$ is the height of the magnetic film.

\subsection{Magnetization model}

The periodic magnetization pattern shown in Fig.~\ref{figSchematic} can be
conveniently expressed as a two-dimensional Fourier series,
\begin{eqnarray}
M_z\!=M_0\!\sum_{n,m=-\infty}^{\infty}\!C_{nm}\cos((n\boldsymbol{k_1}+m\boldsymbol{k_2})\cdot\boldsymbol{r})
\end{eqnarray} with the coefficients \begin{eqnarray}
C_{nm}=\frac{4}{\pi^2}\frac{\sin(\frac{\pi}{2}(m-n))\sin(\frac{\pi}{4}(3m+n))}{(m-n)(3m+n)}
\end{eqnarray}

%\begin{tabular}[t]{l}
 % \multicolumn{2}{r} {\beq

%M_z(x,y)=M_0\sum_{p=1}^{4}\left[\left(\frac{1}{2}+\sum_{n=1}^{n_{max}}\frac{2}{n\pi}\sin\left(\frac{n\pi}{2}\right)\cos\left(n\pi
%(y/a+p/2)\right)\right)\nonumber\\ \times\left(\frac{1}{4}+\sum_{n=1}^{2
%n_{max}}\frac{2}{n\pi}\sin\left(\frac{n\pi}{4}\right)\cos\left(\frac{n\pi}{2}(x/a+p)\right)\right)\right],
%\eeq}\\
% \end{tabular}

%\end{align}
Here, we choose a coordinate system such that the odd Fourier components
vanish. The reciprocal lattice vectors $\boldsymbol{k_1}$ and
$\boldsymbol{k_2}$ are given by $(k_{x},k_{y})=\frac{\pi}{2a}(1,-2)$ and
$\frac{\pi}{2a}(3,2)$ and have length $k_1$ and $k_2$, respectively. We
define $\boldsymbol{r}=(x,y)$ and $a$ is a characteristic length scale of
the lattice geometry as shown in Fig.~\ref{figSchematic}. As the
magnetization is simply the sum of a number of discrete Fourier components
the magnetic field above the surface can be expressed analytically. For
each Fourier component with wavevector $\boldsymbol{k}$ and for $z\gg h$
the magnetic scalar potential is given by: \beq \label{eqPot}
\phi_{\boldsymbol{k}}(r)=\frac{\mu_0}{2}I_{eq}C_{nm}{\rm{e}}^{-kz}\cos(\boldsymbol{k}\cdot\boldsymbol{r}).
\eeq The magnetic field ($\boldsymbol{B}=-\nabla\phi(\boldsymbol{r})$)
above the film surface can be approximated by taking $m,n$=\{-1,0,1\},
leaving two terms:
%\begin{align}
%%%%%
%\begin{widetext}
\beq B_x\!\approx\!\frac{4\mu_0 I_{\rm{eq}}}{
\sqrt{2}\pi^2}\!\left[\frac{k_1{\rm{e}}^{-k_1\! z}}{\sqrt{5}}\!\sin(
{\boldsymbol {k_1}}\!\!\cdot\!\boldsymbol{r})+\frac{k_2{\rm{e}}^{-k_2\!
z}}{\sqrt{13}}\!\sin(\boldsymbol{k_2}\!\cdot\!\boldsymbol{r})\right]
\nonumber \eeq \beq B_y\!\approx\!-\frac{8\mu_0 I_{\rm{eq}}}{
\sqrt{2}\pi^2}\!\left[\frac{k_1{\rm{e}}^{-k_1\!
z}}{\sqrt{5}}\!\sin(\boldsymbol{k_1}\!\!\cdot\!\boldsymbol{r})-\frac{k_2{\rm{e}}^{-k_2\!
z}}{3\sqrt{13}}\!\sin(\boldsymbol{k_2}\!\cdot\!\boldsymbol{r})\right]
\nonumber \eeq \beq B_z\!\approx\!\frac{4\mu_0 I_{\rm{eq}}}{
\sqrt{2}\pi^2}\!\left[k_1{\rm{e}}^{-k_1\!
z}\!\cos(\boldsymbol{k_1}\!\!\cdot\!\boldsymbol{r})+\frac{k_2{\rm{e}}^{-k_2\!
z}}{3}\!\cos(\boldsymbol{k_2}\!\cdot\!\boldsymbol{r})\right] \\\eeq
%\end{widetext}
%\end{align}
where the approximation is valid for distances
$z>\nobreak\frac{1}{k_2}\nobreak\approx\nobreak0.18a$. Each field
component consist of two periodic terms with exponential prefactors which
depend on $z$ and control the respective rate of decay of each Fourier
component from the film surface. For large distances from the film surface
the first term with wave vector $\boldsymbol{k_1}$ dominates and the field
resembles that of a one-dimensional magnetic mirror
potential~\cite{HinHug99}. Closer to the surface the second term involving
$\boldsymbol{k_2}$ introduces additional structure to produce a
two-dimensional periodic potential.

Applying a uniform bias magnetic field oriented along the $y-$direction
partly cancels with the field of the film to create an array of nonzero
field minima.  The applied field strength $B_{0y}$ is related to the trap
height $z_0$ by,

\beq
 B_{0y}=\frac{4\sqrt{2}\mu_0 I_{\rm{eq}}}{\pi^2}\left(\frac{k_1{\rm{e}}^{-k_1z_0}}{\sqrt{5}}+\frac{k_2{\rm{e}}^{-k_2z_0}}{3\sqrt{13}}\right)
\eeq

while the trap positions in $x$ and $y$ are given by symmetry and are
independent of height:
\begin{eqnarray}  &x_{min}=p a\nonumber\\
&y_{min}=(p-1)\frac{a}{2}+2qa
\end{eqnarray}
with the indices $p,q=...-2,-1,0,1,2 ...$.

\begin{figure}
\center
\includegraphics[width=6cm,height=10cm]{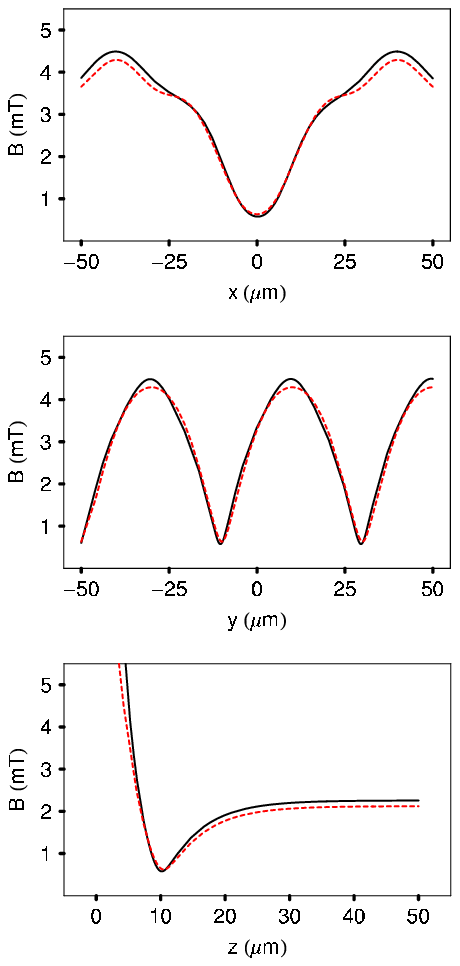}
\caption{(Color online) Magnetic field cross-sections centered on a single
lattice site. The solid lines correspond to an exact numerical calculation
of the magnetic field produced by the magnetic lattice geometry. The
dashed lines are the results of the truncated Fourier series model.}
\label{figCrosssections}
\end{figure}

Our particular design has a length scale $a=20~\mu$m, a magnetization of
$M_0$~=~670~kA/m and a thickness of $h=300$~nm, corresponding to
$I_{\rm{eq}}=0.2$~A. We calculate that a bias field of $B_{0y}=2.25~$mT
results in cigar-shaped Ioffe-Pritchard traps, with their long axis
predominantly oriented in the $x$-direction, positioned 10~$\mu$m from the
film surface~(Fig.~\ref{figCrosssections}). The trapping frequencies are
23~kHz and 6.1~kHz in the radial and axial directions respectively and the
field strength at the trap minimum is 0.57~mT. The barrier heights along
the $\boldsymbol{k_1}$ and $\boldsymbol{k_2}$ directions are 0.4~mK and
1.5~mK, respectively. The trap depth in the $z$-direction is 1.5~mK. The
trapping frequencies can be increased further by increasing $B_{0y}$,
pushing the atoms closer to the chip surface. Tighter radial confinement
can also be achieved by partly cancelling the axial field strength.  The
barrier height between individual lattice sites can be tuned by varying
$B_{0y}$ to change the distance of the traps from the film surface.

\section{{FePt} magnetic films}\label{sec:fept}
Magnetic materials used for experiments with ultracold atoms and
Bose-Einstein condensates should be magnetically hard with a magnetization
sufficient to produce the field strength and gradients for providing tight
confinement to the
atoms~\cite{SinCurHin05,HalWhiSid06,BoyStrPri06,FerGerSpr07}. They should
also have a high coercivity in order to maintain the magnetization when
subjected to externally applied fields. The magnetization should be
oriented perpendicular to the film surface to give maximal design freedom
and should be highly homogeneous to allow for intricate magnetic field
designs and to avoid unwanted spatial variations of the magnetic
field~\cite{WhiHalSid07}. Other desired properties include high corrosion
resistance to facilitate lithographic patterning and good temperature
stability especially during vacuum bake out.

A suitable magnetic material for these experiments is
FePt~\cite{XiaBruBus04,WeiSchFah04,YanPowSel02}. Films of FePt have a high
remanent magnetization \cite{XinBarGerPP}, are extremely stable and
corrosion resistant which allows patterning on the micron scale using
conventional photolithographic techniques. FePt can be created in several
phases. The face centered cubic (fcc) phase is magnetically soft but has a
high saturation magnetization $M_s$=1160~kA/m. The ordered face centered
tetragonal phase (fct), in which Fe and Pt atomic layers are stacked along
the (001) direction, is magnetically hard with high uniaxial anisotropy
constant $K_u=7.7$~MJ/m$^3$.

%These two phases can coexist to form a nanocomposite
%magnet~\cite{LiuLuoSel97} in which the nanocrystalline fct phase couples
%to and aligns the fcc phase. The nanocomposite magnet thus combines
%magnetic hardness with a high saturation magnetization.

\begin{figure}
\center
\includegraphics[width=7cm]{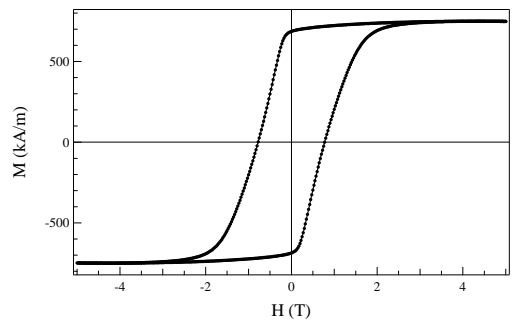}
\caption{Hysteresis curve for a 200~nm thick FePt film. The measured
remanent magnetization of the film is $M_r\approx670$~kA/m and did not
vary significantly for films of different thickness in the range
200-400~nm. } \label{figHitsquid}
\end{figure}

\subsection{Film preparation and characterisation}
We investigated two preparation methods for producing FePt films. The
first set was prepared at the Hitachi San Jose Research Center. The films
are deposited on 300~$\mu$m thick silicon substrates coated with
approximately 10~nm of SiN by magnetron sputter deposition at a
temperature of $T=450^{\circ}$C. The films incorporate a 10~nm thick
amorphous CrTi underlayer and a seed layer of 10~nm RuAl used to directly
grow the FePt in the fct phase in the (001) direction~\cite{SheJudWan05}.
The FePt layer itself was doped with a small amount ($<$10 at.\%) of
carbon to increase the remanent to saturation magnetization ratio.
Evidence for the textured nature of the film was obtained from x-ray
diffraction spectra.

The magnetization of the films were measured by vibrating sample
magnetometry yielding a remanent magnetization of $M_r=$~670~kA/m, a
remanent to saturation magnetization ratio of $M_r/M_s=$~0.93 and a
coercivity of $H_c=$~0.95~T. Figure~\ref{figHitsquid} shows a hysteresis
curve of a 200~nm thick film. There were no significant differences
between films of various thickness in the range 200-400~nm.

To test the suitability of our FePt films for atom optics applications we
magnetize and then bake the deposited films for 3 hours at 150~$^{\circ}$C
in air to characterize temperature stability. After baking the
magnetization had reduced by $\sim$~3~$\%$. The films were also tested
against demagnetization due to time varying magnetic fields. An
experimental cycle typically involves external fields of up to 20~mT with
a cycle time of about 30~s. To test the stability of the magnetization in
those fields the films were magnetized and minor hysteresis loops were
taken over the range of -20~mT to 20~mT. The magnetization did not
measurably decrease after 15 cycles. Additional long-term tests performed
within the experimental apparatus also indicate the magnetization is
stable during typical experimental cycles.

We have also measured the surface roughness and grain size of the films
using atomic force microscopy, scanning electron microscopy and x-ray
diffraction. Surface roughness can be a significant source of spatial
magnetic field noise. The {\it{rms}} height variation in the film is
$\sim$~6~nm. The grain size is related to the width of the x-ray
diffraction peaks via the Debye Scherrer equation. This yields an average
grain size of $\sim$~35~nm. Taking these numbers into account we
anticipate spatial magnetic field variations around 0.4~$\mu$T which
corresponds to energy variations between lattice sites of less than the
vibrational level spacing ($<$~6~kHz)~\cite{WhiHalSid07}.

As a second route to produce hard magnetic FePt films we investigated the
possibility of rapid post annealing of films magnetron sputtered directly
onto silicon at lower temperatures (250$^\circ$C). After deposition at
this temperature the FePt is in a highly disordered partly amorphous fcc
soft magnetic phase. The film is annealed at a temperature of 450$^\circ$C
for 3~minutes in order to initiate the hard magnetic fct phase. This film
had a remanent magnetization of $M_r=$~580~kA/m, a remanent to saturation
magnetization ratio of $M_r/M_s=$~0.80 and a coercivity of $H_c=$~0.95~T.
Although the magnetization is somewhat smaller than for the first film the
post annealed film has the advantage that it has a smaller surface
roughness (1~nm {\it rms} height variation) and smaller grains
($\sim$~20~nm diameter). Furthermore, the annealed film is straightforward
to produce without the complication of underlayers. It was however less
stable against elevated temperatures and showed a 20~\% decrease in
magnetization after a 3 hour 150~$^{\circ}$C bake, persuading us to use
the Hitachi film for the experiment.

\subsection{Patterning}

\begin{figure}
\center
\includegraphics[width=7cm]{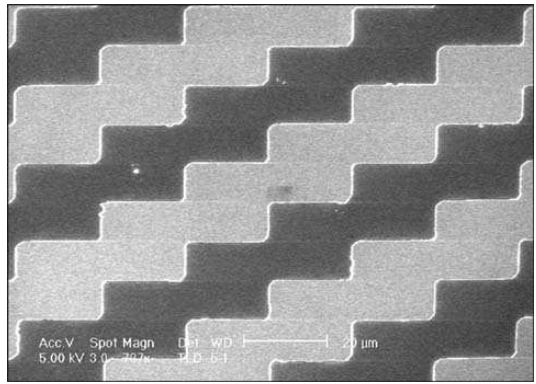}
\caption{ Scanning electron microscopy image of the permanent magnetic
array. The light gray areas correspond to the FePt pattern and the darker
regions are the Si substrate. } \label{figSEM}
\end{figure}

To pattern our films we use optical lithography. We used a commercially
obtained custom designed Cr mask. This mask was patterned with a 442~nm
HeCd laser with a resolution of $\sim1~\mu$m.  The deposited films are
first spin coated with $\sim1~\mu$m optical resist and then illuminated
for 4~s using a I-line (365 nm) mask aligner. After developing the samples
for one minute the resist on the illuminated parts of the sample is
dissolved. The surface can be cleaned with a low power, two minute oxygen
etch using a dry etching system. The exposed FePt surface was then etched
using an argon plasma for about 30~minutes. The remaining resist covering
the chip structures was then removed in an ultrasonic bath. After this the
chip was investigated using optical and scanning electron microscopy
(SEM). A SEM image of the lattice is shown in Fig.~\ref{figSEM}. Finally,
the magnetic films were coated with a 100~nm reflective gold over layer
and were magnetized in a 5~T magnetic field.

\subsection{Chip construction}

\begin{figure}
\center
\includegraphics[width=6cm]{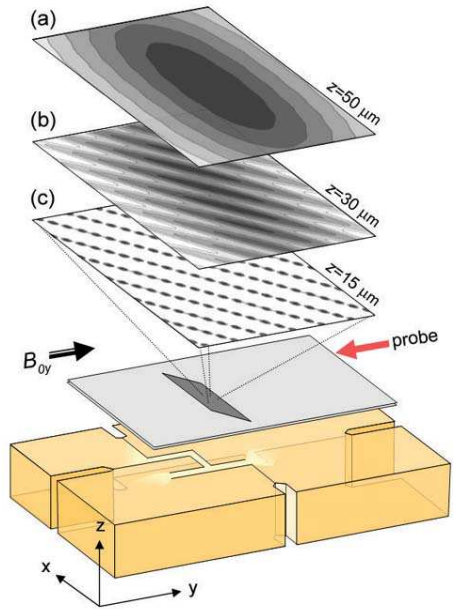}
\caption{ (Color online) Layout of the full atom chip, showing the
permanent magnetic chip with the lattice and the wire structure used for
loading the lattice. Also shown are magnetic field contour plots during
various stages of the loading process. In (a) the atoms are trapped by the
Z-shaped wire with an external bias field $B_{\rm{0y}}$. In (b) the
current in the Z-wire is turned down, moving the cloud closer to the
surface and the potential gets corrugated by the lattice. In (c) the
current is completely turned off and the atoms are trapped in the lattice
sites. We also show the orientation of the probe beam used to make the
absorption images shown in Fig~\ref{figlattice}.} \label{figChipmount}
\end{figure}

The completed atom chip consists of the silicon substrate with the
patterned FePt film epoxied to a copper mount which incorporates a wire
structure used to load the microtraps~(Fig.~\ref{figChipmount}). In
particular, the atom chip includes (besides the FePt film) a U-shaped
wire, a Z-shaped wire and a small rf coil (ten turns, diameter 5~mm) used
to apply radio frequency fields for forced evaporative cooling. The chip
wires are $300~\mu$m wide, have a height of $\sim100~\mu$m and are capable
of running up to 15~A continuously without significant heating. The atom
chip is mounted face-down in a glass cell vacuum chamber, surrounded by a
system of three orthogonal pairs of coils for producing external
fields~\cite{FerGerSpr07}.

\section{Loading atoms to the lattice potential}
\label{secLoading}

Laser cooled $^{87}$Rb atoms are transferred to the atom chip using
techniques common to many atom chip experiments. After an initial cooling
stage the atoms can be transferred to the magnetic lattice potential. The
following section describes the loading process.

\subsection{Magnetic trapping and evaporative cooling}

We collect approximately $1\times10^8$ $^{87}$Rb atoms in a mirror magneto
optical trap (mMOT) 4~mm from the chip surface. This mMOT relies on a
quadrupole magnetic field (0.1~T/m) produced by a pair of external coils
overlapped with four laser beams with appropriate detuning and
polarizations.

The atoms are transferred closer to the chip by passing 5.5~A through the
U-shaped wire, applying a bias field of $B_{0y}=0.44$~mT, and ramping off
the external magnetic quadrupole field. This U-MOT has a field gradient of
0.2~T/m. To further cool the atoms a 4~ms polarisation gradient cooling
stage is applied by detuning the trap laser below resonance by 36~MHz and
rapidly reducing the U-wire current and bias field by a factor of three to
weaken the field gradient.  We found that this stage significantly
improves the loading efficiency into the magnetic trap. To magnetically
trap the atoms a 0.2~ms optical pumping pulse in a uniform magnetic field
drives the atoms to the $|F=2,m_F=+2\rangle$ state. Then the U-wire
current is reduced to zero, the Z-wire current is increased to 12~A and
the bias field $B_{0y}$ is increased to 1.4~mT to produce an
Ioffe-Pritchard magnetic trap containing $2\times10^7$ atoms. We then
adiabatically compress the trap by increasing the Z-wire current to 15~A
and by increasing $B_{0y}$ to 4.0~mT. The calculated radial and axial trap
frequencies are $2\pi\times 630$~Hz and $2\pi\times 27$~Hz respectively.
The mean elastic collision rate is $100$~s$^{-1}$.

Forced RF evaporative cooling is applied by a programmable DDS rf
synthesiser via a 2~W rf amplifier to the built-in rf coil. The
evaporation stage consists of three linear rf sweeps: from 15~MHz to 6~MHz
in 2.5~s, from 6~MHz to 3~MHz in 1.5~s and from 3~MHz to the final
evaporation frequency of 520~kHz in a further 2~s. To improve the
evaporation efficiency we also compress the Z-wire trap further after the
first sweep by increasing $B_{0y}$ to 5.0~mT which pushes the trap closer
to the surface and increases the radial trap frequency to $2\pi\times
1400$~Hz. For a final radio frequency of 545~kHz the cloud temperature
reaches the Bose-Einstein condensation (BEC) temperature
($T_c\approx0.5~\mu$K) with approximately $3\times10^4$~atoms. The BEC
transition is accompanied by the onset of a bimodal density distribution
and anisotropy in the time of flight expansion. Continuing the rf sweep to
520~kHz we produce an almost pure BEC containing 6000~atoms.

\subsection{Transfer to the lattice}
To load atoms to the lattice we truncate the rf sweep at a final frequency
of 800~kHz to produce a cold thermal cloud of $1.5\times10^5$~atoms at an
estimated temperature of $3~\mu$K. We load the lattice with a thermal
cloud rather than a Bose-Einstein condensate to ensure each lattice site
contains a sufficient number of atoms for detection and to minimise the
influence of heating during loading.  We expect that further rf
evaporation could be performed after loading with high efficiency due to
the extremely tight confinement in individual lattice sites. Loading the
lattice is rather straightforward as the magnetic field produced by each
lattice site is by design oriented in the same direction as that produced
by the Z-shaped wire.  In this way we can simply reduce the Z-wire current
to zero, pushing the trapped atoms closer to the surface and finally into
the lattice potential.   The atoms are imaged using a resonant probe laser
aligned along $y$ with a small inclination ($\sim1^\circ$) with respect to
the chip surface to produce a reflection image. Reflection imaging allows
us to image atom clouds in close proximity to the chip surface and to
determine the height of the cloud. One disadvantage however is that the
absorption signal is integrated along the $y-$direction.

\begin{figure}
\center
\includegraphics[width=8cm]{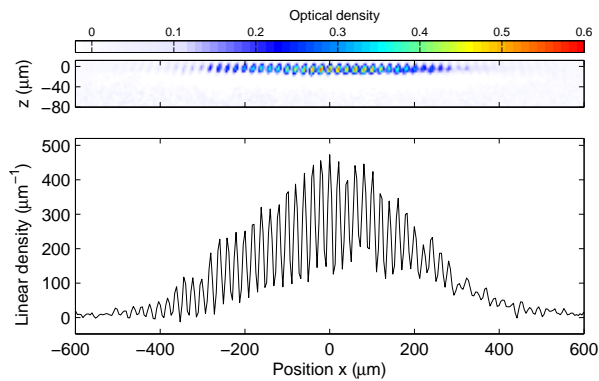}
\caption{ (Color online) Absorption image of the loaded array (top) and
vertically integrated atom density. The periodicity of 20~$\mu$m due to
the array sites can be clearly seen. The contrast between individual sites
may be limited by the integration along the field of view. }
\label{figlattice}
\end{figure}

In the experiment we reduce the Z-wire current from 15~A to zero over
50~ms to move the trap closer to the FePt film. For a Z-wire current of
2~A the cloud expands in the axial direction, moves to approximately
$20~\mu$m from the FePt surface and becomes clearly corrugated by the
lattice potential, as is schematically shown in Fig.~\ref{figChipmount}.
Further decreasing the Z-wire current to zero transfers all of the atoms
to the magnetic lattice (Fig.~\ref{figlattice}).  The spacing between
lattice sites determined from an absorption image of the cloud is
$20~\mu$m, as expected from the lattice geometry. The overall gaussian
density distribution is due to the initial distribution of atoms in the
Z-wire trap and does not expand further after loading. We observe a
minimum of 30 occupied lattice sites spanning $600~\mu$m along $x$ but the
true number of occupied sites may be higher. A better estimate of the
number of occupied lattice sites may be obtained in the future by
inclining the probe beam further to resolve the lattice structure along
the $y-$axis. Assuming adiabatic compression during loading we expect the
individual atom clouds to be localized, each at a temperature of
$\sim100~\mu$K, smaller than the inter-well barrier height and trap depth
($\sim0.5$~mK). We observe a lifetime of more than 1~s for the atoms
trapped in the lattice potential.

\section{Discussion}

Our present observation shows clouds of trapped atoms spaced by 20~$\mu$m
along the $x$-direction. It is likely that at least a few sites are
occupied along the detection direction. Large arrays of microtraps can
serve as a quantum register and a shift
register~\cite{DotAltRau05,BeuTucGra07}. Field calculations show that we
should be able to shift the traps in the direction perpendicular to
$\boldsymbol{k_2}$ by applying uniform external $B$-fields
\cite{GerSpr06,HinHug99}.

The present density of lattice sites is 1250 traps/mm$^2$. It should be
feasible to scale down the linear dimensions of our structure into the
(sub-)micron range and thus increase the trap density by two or three
orders of magnitude. We are presently investigating the possibilities to
drive single and two-qubit gate operations \cite{CalHinZol00}. We plan to
directly address individual sites by a combination of static and radio
frequency or microwave fields, to selectively change the internal state of
the atoms. An important parameter for the optical manipulation of the
qubits is the confinement of the atoms on the scale of the optical
wavelength. The present confinement is already sufficiently tight for the
trap ground state to be in the so-called Lamb-Dicke limit, where the
photon recoil energy is less than the vibrational level splitting. Further
downscaling in size should make the confinement even tighter, and thus
decrease the Lamb-Dicke parameter. We plan to implement rf evaporative
cooling in the lattice to produce ground state atoms in every lattice
site. Together these implementations will be important first steps towards
future quantum information processing applications on atom chips
\cite{TreSteHan06}.

To summarize, we described the design and production of a permanent
magnetic array of Ioffe-Pritchard microtraps for cold atoms based on FePt.
We have given analytical expressions for the magnetic field produced by
this lattice. We have investigated magnetic films for producing the
lattice. We have loaded more than 30 lattice sites with cold atoms. The
loading protocol relies on current carrying structures beneath the
permanent magnetic chip. The lifetime of the atoms trapped in the lattice
exceeds 1~s.

\begin{acknowledgements}

We gratefully acknowledge technical support from Chris R{\'e}tif and Huib
Luigjes. This work is part of the research program of the Stichting voor
Fundamenteel Onderzoek van de Materie (Foundation for the Fundamental
Research on Matter) and was made possible by financial support from the
Nederlandse Organisatie voor Wetenschappelijk Onderzoek (Netherlands
Organization for the Advancement of Research) and by the fabrication and
characterization facilities of the Amsterdam {\it nano}Center. It was also
supported by the EU under contract MRTN-CT-2003-505032.

\end{acknowledgements}

\end{document}